\def\la{\hbox{{\lower -2.5pt\hbox{$<$}}\hskip -8pt\raise
-2.5pt\hbox{$\sim$}}}
\def\ga{\hbox{{\lower -2.5pt\hbox{$>$}}\hskip -8pt\raise
-2.5pt\hbox{$\sim$}}}
\def\ltsima{$\; \buildrel < \over \sim \;$}
\def\simlt{\lower.5ex\hbox{\ltsima}}
\def\gtsima{$\; \buildrel > \over \sim \;$}
\def\simgt{\lower.5ex\hbox{\gtsima}}

\documentstyle[epsfig]{elsart}

\begin{document}
\begin{frontmatter}
\title{The gamma ray background from large scale structure formation}
\author[UniFI]{Stefano Gabici\thanksref{corr1}}
\address[UniFI]{Dipartimento di Astronomia e Scienza dello Spazio, Universit\`a di
Firenze,\\ Largo E. Fermi, 5 I-50125 Firenze (Italy)}
\thanks[corr1]{E-mail: gabici@arcetri.astro.it}
\author[inaf]{Pasquale Blasi\thanksref{corr2}}
\address[inaf]{INAF/Osservatorio Astrofisico di Arcetri,\\
Largo E. Fermi, 5 I-50125 Firenze (Italy)}
\thanks[corr2]{E-mail: blasi@arcetri.astro.it}

\begin{abstract}
Hierarchical clustering of dark matter halos is thought to describe well
the large scale structure of the universe. The baryonic component 
of the halos is shock heated to the virial temperature while a small 
fraction of the energy flux through the shocks may be energized 
through the first order Fermi process to relativistic energy per 
particle. It has been proposed that the electrons accelerated in this 
way may upscatter the photons of the universal microwave background 
to gamma ray energies and indeed generate a diffuse background of 
gamma rays that compares well to the observations.
In this paper we calculate the spectra of the particles accelerated
at the merger shocks and re-evaluate the contribution of structure
formation to the extragalactic diffuse gamma ray background (EDGRB),
concluding that this contribution adds up to at most $10\%$ of the
observed EDGRB. 
\end{abstract}

\end{frontmatter}

\section{Introduction}

EGRET observations \cite{EGRET} showed that the universe is permeated by 
a background of gamma radiation that seems to exceed the flux of gamma rays 
expected from cosmic ray interactions in our own Galaxy, as calculated using
theoretical models of the origin and propagation of cosmic rays. 
This excess has long been considered of extragalactic 
origin, and innumerable attempts to explain it in terms of different kind
of sources have been made. Whether this radiation is the result of discrete 
unresolved extragalactic sources or rather a truly diffuse background is 
still unknown, and is matter of investigation for future gamma ray telescopes 
such as GLAST. 

It is somehow disturbing that the extragalactic origin for this 
background has been inferred from a combination of measurements and 
theoretical modelling of the diffuse galactic gamma radiation. In fact
some authors \cite{dar} have proposed that observations can be also 
explained as a result of a population of galactic relativistic electrons 
upscattering the microwave and starlight radiations to gamma ray energies 
through inverse Compton scattering (ICS). These electrons would not be 
correctly accounted for in standard models of cosmic ray propagation in 
the Galaxy. 

If this radiation is in fact mainly extragalactic, its source or sources 
need to be found. While it is believed that blazars may contribute a large
fraction of the extragalactic diffuse gamma ray background \cite{setti,ciccio,blone,ss}
(EDGRB), it is not yet clear whether they can saturate it (see, for example \cite{bltwo,blthree}). Around 1 GeV a non negligible contribution to the diffuse background might also come from normal galaxies \cite{bho}.

In the last few years, clusters of galaxies have been proposed as
sources of high energy gamma rays and in fact even as sources of the 
EDGRB. The first paper to propose this possibility is Ref. \cite{ds95}.
Some problems were identified in these calculations and discussed in Refs.
\cite{stecker,bbp97}. A detailed calculation of the EDGRB due to clusters
of galaxies was carried out in \cite{bc98}, where the authors concluded 
that not more than a few percent of the observed background of gamma
rays could be accounted for in terms of hadronic interactions in clusters
of galaxies. 

More recently, the authors in Ref. \cite{lw,totani} have reproposed a connection
between the EDGRB and clusters of galaxies. More correctly the connection 
should exist between the EDGRB and the process of hierarchical large scale 
structure formation. The claim is that the whole EDGRB can be explained in terms of
ICS of electrons accelerated at shocks during structure formation up to 
ultrarelativistic energies. Shocks form naturally during the merger of
two halos that generate a new bigger structure.

In a recent paper \cite{gb2002} we have studied in detail the process of 
acceleration and reacceleration of particles at shocks during mergers of
clusters of galaxies, and we have proposed a semi-analytical method to 
evaluate in a self-consistent manner the Mach numbers of the shocks developed 
at each merger event. 
The Mach numbers are related in a unique way to the strength of the
shocks and therefore to their ability to accelerate particles. We adopt 
here the same method introduced there and apply it to the calculation of the 
contribution of structure formation to the EDGRB. 

Our conclusion is that at most $10\%$ of the observed EDGRB can be 
explained by invoking the process described above. We discuss in detail the 
reasons for the difference between our results and those in \cite{lw,totani}.

The paper is planned as follows: in \S \ref{sec:formation} we describe 
the basics of hierarchical clustering and the formation of shock surfaces
during structure formation. In \S \ref{sec:acceleration} we summarize 
the physics of shock acceleration and specialize the discussion to the
case of merger shocks. In \S \ref{sec:diffuse} we illustrate our calculations
of the diffuse gamma ray background from structure formation. We conclude
in \S \ref{sec:conclusions}.

\section{Structure formation and related shocks}\label{sec:formation}

The standard theory of structure formation predicts that larger structures 
result from the mergers of smaller structures, which on average are formed at 
earlier times.
Press and Shechter \cite{PS74} (hereafter PS) were the first to propose an 
efficient analytical description of the hierarchical clustering. It represents 
an extremely powerful tool that allows one to reconstruct realizations 
of the merger history of a cluster with fixed mass at the present time. 
There are now different flavors of these analytical methods with different
levels of sophistication \cite{bond91,ravi}.

In \cite{gb2002} we described in detail the procedure adopted to simulate 
the merger history of a cluster. We summarize here the basic points involved
in this procedure.
 
In the PS formalism, the differential comoving number density of clusters with 
mass $M$ at cosmic time $t$ can be written as:
\begin{equation}
\frac{dn(M,t)}{dM}=\sqrt{\frac{2}{\pi}}\,\frac{\varrho}{M^2}\,
\frac{\delta_c(t)}
{\sigma(M)}\,\left|{\frac{d\ln \sigma(M)}{d\ln M}}\right| exp\left[-\frac
{\delta_c^2(t)}{2\sigma^2(M)}\right].
\end{equation}
The rate at which clusters of mass $M$ merge at a given time $t$ is written 
as a function of $t$ and of the final mass $M^{\prime}$ \cite{LC93}:
$${\cal R}(M,M^{\prime},t)dM^{\prime}=$$
\begin{eqnarray}
\sqrt{\frac{2}{\pi}}\,\left|
\frac{d\delta_c(t)}{dt}\right|\,\frac{1}{\sigma^2(M^{\prime})}\,
\left|\frac{d\sigma(M^{\prime})}{dM^{\prime}}\right|\,
\left(1-\frac{\sigma^2(M^{\prime})}{\sigma^2(M)}\right)^{-3/2} \nonumber \\
\rm{exp}\left[-\frac{\delta_c^2(t)}{2}\left(\frac{1}{\sigma^2(M^{\prime})}-
\frac{1}{\sigma^2(M)}\right)\right]dM^{\prime},
\end{eqnarray}
where $\varrho$ is the present mean density of the universe, $\delta_c(t)$ 
is the critical density contrast linearly extrapolated to the present time 
for a region that collapses at time $t$, and $\sigma(M)$ is the current rms 
density fluctuation smoothed over the mass scale $M$.
For $\sigma(M)$ we use an approximate formula proposed in \cite{kita}, 
normalized by assuming a bias parameter $b=0.9$. We adopt the expression of
$\delta_c(t)$ given in \cite{NS}. In this respect our approach
is similar to that adopted in \cite{fujita}.

\begin{figure}[thb]
 \begin{center}
  \mbox{\epsfig{file=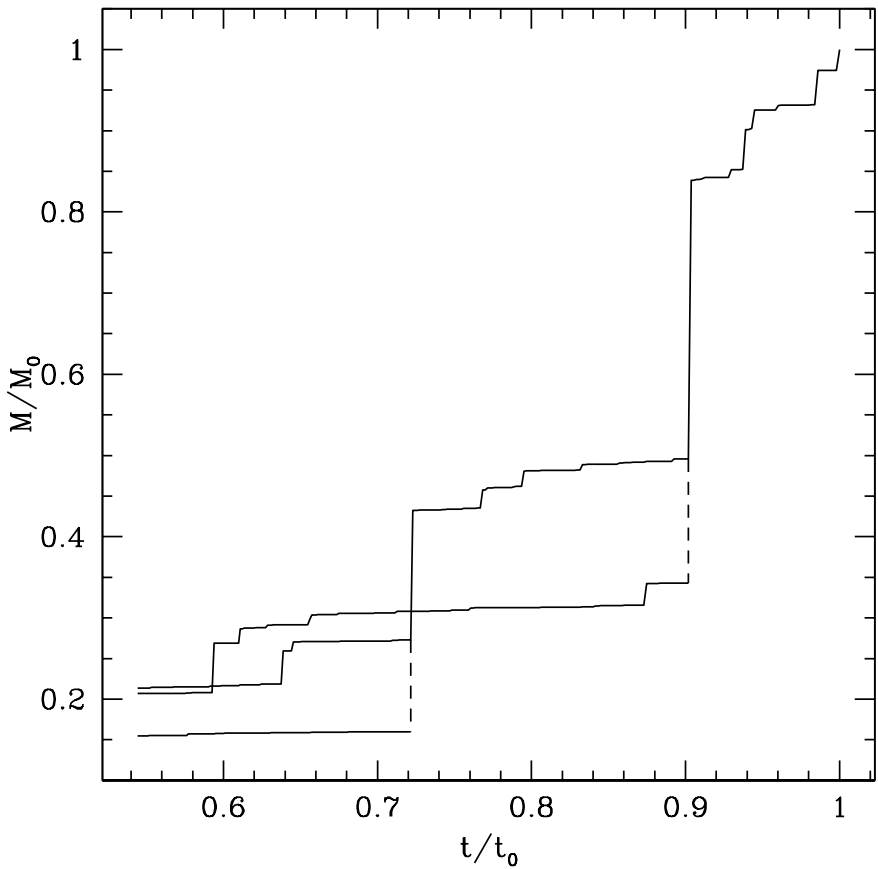,width=13.cm}}
  \caption{\em {Merger history of a cluster with present mass $10^{15}$ 
solar masses. The mass (y-axis) suffers major jumps in big merger events. 
Time is on the x-axis.}
\label{fig:tree}}
 \end{center}
\end{figure}

In fig. \ref{fig:tree} we plot a possible realization of the merger 
tree for a cluster with present mass of $10^{15} M_\odot$. 
The history has been followed back in time up to redshift 
$z=3$. The big jumps in the cluster mass correspond to merger events, while 
smaller jumps correspond to what in the literature are known as accretion 
events. 
During the merger of two clusters of galaxies, the baryonic component,
feeling the gravitational potential created mainly by the dark matter 
in the two cluster, is forced to move supersonically and shock waves are 
generated in the intracluster medium.

In order to properly describe the physical properties of these shocks, we 
use here the approach introduced in \cite{Takizawa99,gb2002}. 
We assume to have two dark matter halos, as completely 
virialized structures, at temperatures $T_1$ and $T_2$, and with masses 
$M_1$ and $M_2$ (here the masses are the total masses, dominated by the dark 
matter components). The virial radius of each cluster can be written as follows
\begin{equation}
r_{vir,i} = \left(\frac{3 M_i}{4 \pi \Delta_c \Omega_m \rho_{cr} (1+z_{f,i})^3}
\right)^{\frac{1}{3}}=
\left(\frac{G M_i}{100 \Omega_m H_0^2 (1+z_{f,i})^3}\right)^{\frac{1}{3}},
\label{eq:vir}
\end{equation}
where $i=1,2$, $\rho_{cr}=1.88\times 10^{-29} h^2 \rm{g}~\rm{cm}^{-3}$ 
is the current value of the critical density of the universe, 
$z_{f,i}$ is the redshift of formation of the halo $i$, $\Delta_c=200$ is 
the density constrast for the formation of the halo and $\Omega_m$ is the 
matter density fraction. 
In the right hand side of the equation we used the fact 
that $\rho_{cr}=3 H_0^2/8\pi G$, where $H_0$ is the Hubble constant.
The formation redshift $z_f$ is on average a decreasing function of the mass,
meaning that smaller structures form at larger redshifts, consistently
with the hierarchical scenario of structure formation. There are intrinsic
fluctuations in the value of $z_f$ at fixed mass, due to the stochastic 
nature of the merger tree. The formation redshift $z_f$ is calculated here
following \cite{LC93}\footnote{We adopt here as formation redshift the
peak value of the distribution given in \cite{LC93}.}.

The relative velocity of the two merging structures, $V_r$, can be easily 
calculated from energy conservation:
\begin{equation}
-\frac{G M_1 M_2}{r_{vir,1}+r_{vir,2}} + \frac{1}{2} M_r V_r^2 = 
-\frac{G M_1 M_2}{2 R_{12}},
\label{eq:Vrel}
\end{equation}
where $M_r=M_1 M_2 / (M_1+M_2)$ is the reduced mass and $R_{12}$ is the 
turnaround radius of the system, where the two subhalos are supposed to
have zero relative velocity. In fact the final value of the relative velocity
at the merger is quite insensitive to the exact initial condition of the two
subclusters. In an Einstein-De Sitter cosmology this spatial scale equals 
twice the virial radius of the system. Therefore, using eq. (\ref{eq:vir}), 
we get:
\begin{equation}
R_{12} = 2\,\left( \frac{M_1+M_2}{M_1}\right)^{1/3} \frac{1+z_{f,1}}{1+z_f} 
r_{vir,1}.
\end{equation}
where $z_f$ is the formation redshift of the halo with mass $M_1+M_2$.
This expression remains valid in approximate way also for other cosmological
models \cite{Lahav91}. The sound speed of the halo $i$ is given by 
$$
c_{s,i}^2 = \gamma_g (\gamma_g - 1) \frac{G M_i}{2 r_{vir,i}},
$$
where we used the virial theorem to relate the gas temperature to the 
mass and virial radius. The adiabatic index of the gas is
$\gamma_g=5/3$. The Mach number of each cluster while moving in the potential
of both clusters can be written as follows:
\begin{eqnarray}
{\cal{M}}^2_1 & = & \frac{4(1+\eta)}{\gamma(\gamma-1)}
\left[\frac{1}{1+\frac{1+z_{f,1}}
{1+z_{f,2}}\eta^{1/3}}-\frac{1}{4\frac{1+z_{f,1}}{1+z_{f}}(1+\eta)^{1/3}}
\right] \nonumber \\
{\cal{M}}^2_2 & = & \eta^{-2/3}\frac{1+z_{f,1}}{1+z_{f,2}}{\cal{M}}^2_1, 
\end{eqnarray}
where $\eta=M_2/M_1<1$.
The procedure illustrated above can be applied to a generic couple of 
merging halos, and in particular it can be applied to a generic merger
event in the history of a cluster with fixed mass at the present time.

\begin{figure}[thb]
 \begin{center}
  \mbox{\epsfig{file=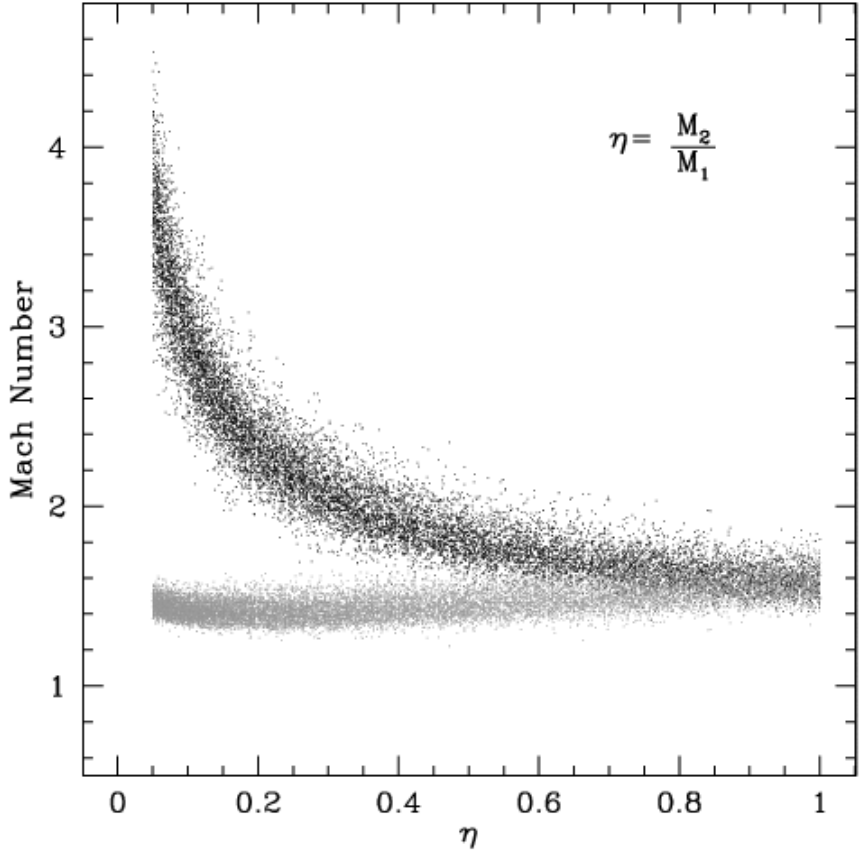,width=13.cm}}
  \caption{\em {Distribution of the Mach numbers of 
merger related shocks as a function of the mass ratio of the merging
subclusters. The upper strip is the distribution of Mach numbers in
the smaller cluster, while the lower strip refers to the bigger cluster.}
\label{fig:mach}}
 \end{center}
\end{figure}

The results of our calculations of 500 realizations of the merger history produce 
the Mach numbers plotted in fig. \ref{fig:mach}. 
The striking feature of this plot is that for major mergers, involving 
clusters with comparable masses ($\eta\sim 1$), the Mach numbers of the 
shocks are of order unity. In other words the shocks are only moderately 
supersonic. 
In order to achieve Mach numbers of order of $3-4$ it is needed to consider 
mergers between clusters with very different masses ($\eta\sim 0.05$).
These events are the only ones that produce strong shocks, and this is of crucial 
importance for the acceleration of suprathermal particles, and, as discussed below,
for the calculation of the spectrum of the diffuse gamma rays generated by the
accelerated particles. 

Each merger here is assumed to be a two body event, namely the potential well
is assumed to be dominated by the two merging structures. It may be argued that
the merger between two objects may occur in the deeper gravitational potential 
created by nearby structures. In this case, the relative velocity between the 
two clusters, and also the related shock Mach numbers  may be larger (or smaller) 
than those estimated above.
Following \cite{gb2002} we briefly discuss here a simple argument that suggests
that this problem should not be relevant for our purposes.
We assume that the two clusters, with mass $M_1$ and $M_2$, are merging in a volume 
of average size $R_{sm}$ where the smoothed overdensity is $1+\delta$ ($\delta=0$ corresponds 
to matter density equal to the mean value). Clearly the overdense region must contain 
more mass than that associated with the two halos, therefore for a top-hat 
overdensity at $z=0$ we can write:
\begin{equation}
\frac{4}{3} \pi R_{sm}^3 \rho_{cr} \Omega_m (1+\delta) = \xi (M_1+M_2),
\end{equation}
where $\xi>1$ is a measure of the mass in the overdense region in excess of 
$M_1+M_2$. In numbers, using $\Omega_m=0.3$, this condition becomes:
\begin{equation}
(1+\delta) = 2 \xi M_{15} R_{10}^{-3},
\label{eq:delta}
\end{equation}
where $M_{15}$ is $M_1+M_2$ in units of $10^{15}$ solar masses and
$R_{sm}=10~{\rm Mpc}~R_{10}~h^{-1}$. 

If the clusters are affected by the potential well of an overdense region with
total mass $M_{tot}$, the maximum relative speed that they can acquire is 
$v_{max}\approx 2 \sqrt{G M_{tot}/R_{sm}}$. Note that this would be the relative 
speed of the two clusters if they merged at the center of the overdense region and 
with a head-on collision, therefore any other (more likely) configuration would 
imply a relative velocity smaller than $v_{max}$. In particular, the presence of
the local overdensity might even cause the two merging clusters to slow down,
rather than a larger relative velocity. 
In numbers 
$$ v_{max} =1.1\times 10^8 \xi^{1/2} M_{15}^{1/2} R_{10}^{-1/2} ~
{\rm cm/s}.$$
Using the usual expression for the sound speed in a cluster with mass $M_i$ we also
get
$$c_s = 8.8 \times 10^7 M_{i,15}^{1/3}~ {\rm cm/s}.$$
Therefore the maximum Mach number that can be achieved in the i-th cluster is
\begin{equation}
{\cal M}_{i,max} = 1.25 \xi^{1/2} R_{10}^{-1/2} M_{15}^{1/2} M_{i,15}^{-1/3}.
\end{equation}
As stressed in the previous sections, the Mach numbers which may be relevant for
particle acceleration are ${\cal M}>3$, which implies the following condition on 
$\xi$:
\begin{equation}
\xi>5.8 M_{15}^{-1} M_{i,15}^{2/3}R_{10},
\end{equation}
that, when introduced in eq. (\ref{eq:delta}) gives:
\begin{equation}
(1+\delta) > 11.6 R_{10}^{-2} M_{i,15}^{2/3}.
\label{eq:overdensity}
\end{equation}
Similar results may be obtained using the velocity distribution of dark matter 
halos as calculated in semi-analytical models \cite{antonaldo} and 
transforming this distribution into a pairwise velocity distribution, by adopting
a suitable recipe.

The probability to have an overdensity $1+\delta$ in a region of size $R_{sm}$
has the functional shape of a log-normal distribution, as calculated in \cite{lognormal}. 
Eq. (\ref{eq:overdensity}) gives the overdensity 
$1+\delta$ necessary for a cluster of mass $M_i$ to achieve a Mach number at 
least 3 in the collision with another cluster in the same overdense region. 
The probabilities $P(\delta)$ as a function of the mass of the cluster $M_i$
were estimated in \cite{gb2002}. For rich clusters, with masses larger than 
$5\times 10^{14}M_\odot$ (corresponding to X-ray luminosities 
$L_X>4\times 10^{44} erg/s$) the probability that the presence of a local overdensity 
may generate Mach numbers larger than $3$ has been estimated to be less than 
$10^{-3}$, suggesting that our two body approximation is reasonable, in particular
for the massive clusters that are typically observed to have nonthermal activity. 
For smaller clusters, the probabilities become higher, indicating that the 
distribution of Mach numbers might have a larger spread compared with that 
illustrated in fig. \ref{fig:mach}. 
Note however that for small clusters, even the two body approximation gives 
relatively high Mach numbers, provided the merger occurs with a bigger cluster,
simply as a result of a lower temperature and a correspondingly lower sound speed. 

Motivated by these arguments, in the following we keep the assumption of 
binary mergers.

\section{Shock acceleration during structure formation}\label{sec:acceleration}

Merger related shocks may serve as cosmic ray accelerators. In this section 
we estimate the maximum energy achievable in one of these shocks for acceleration
of electrons. We aim at using the most optimistic situation, to achieve the 
highest maximum energy for the accelerated particles. For this reason, we adopt
a Bohm diffusion coefficient
$$D(E) = \frac{1}{3} r_L c = 3.3\times 10^{22} E(GeV) B_\mu^{-1} \rm cm^2/s,$$
where $E$ is the particle energy in GeV and $B_\mu$ is the magnetic field at
the shock in units of $\mu G$. Other possible choices for the diffusion coefficent
are discussed in \cite{blasi 2001,gb2002}.

The acceleration time is defined as follows:
\begin{equation}
\tau_{acc} = \frac{3}{u_1-u_2} \left[\frac{D_1}{u_1} + \frac{D_2}{u_2}\right]\approx
\frac{3 D(E)}{u_1^2} \frac{r(r+1)}{r-1},
\end{equation}
where in the last step we neglected the jump in the magentic field at the shock,
and we introduced the ratio $r=u_1/u_2$ of the two velocities upstream and 
downstream the shock. The compression ratio $r$ is related to the Mach number 
of the shock through the relation 
$$r=\frac{\frac{8}{3} {\cal M}^2}{\frac{2}{3} {\cal M}^2 +2}.$$
For relativistic electrons the main channel of energy losses is represented by
ICS on the photons of the microwave background. The time scale for these losses
is $$\tau_{ICS}\approx 4\times 10^{16} E(GeV)^{-1}~s.$$

Requiring that acceleration occurs faster than losses implies the following 
maximum energy
\begin{equation}
E_{max} = 6.3 \times 10^4 B_\mu^{1/2} h(r)^{-1/2} u_8 ~ GeV,
\label{eq:Emax}
\end{equation}
where $h(r)=r(r+1)/(r-1)$ and $u_8$ is the collision speed in units of $10^8$ cm/s.
It is worth stressing that larger diffusion coefficients would significantly reduce 
the value of the maximum energy compared to that given in Eq. (\ref{eq:Emax}). This
effect is discussed in \cite{blasi 2001} and \cite{gb2002}.

It is instructive to evaluate the energy of the photons radiated by electrons
with energy $E_{max}$ due to synchrotron emission and ICS respectively. The peak
frequency for synchrotron emission is  
$$\nu_{max}=1.4\times 10^{16} B_\mu^2 h(r)^{-1} u_8^2 ~~ Hz.$$
This corresponds to an energy 
$$h \nu_{max}\approx 58 B_\mu^2 h(r)^{-1} u_8^2 ~~ eV.$$
The peak energy for ICS off the microwave photons is 
$$\epsilon_{ICS} \approx 10 B_\mu h(r)^{-1} u_8^2 ~~ TeV.$$
The choice of a diffusion coefficient that is larger than the Bohm coefficient
would considerably decrease the values of $\nu_{max}$ and $\epsilon_{ICS}$. 
The gamma ray production from ICS strongly relies upon the assumption of
Bohm diffusion.

The spectrum of the accelerated electrons is provided by the standard theory
of first order shock acceleration, and in the linear regime is uniquely 
determined by the compression factor (or equivalently the Mach number) of
the shock. In fact the spectrum can be written as a power law $E^{-\gamma}$
with $\gamma=(r+2)/(r-1)$. The slope tends asymptotically to $\gamma=2$ for 
strong shocks ($r\sim 4$).

\section{The diffuse gamma ray background from hierarchical clustering}
\label{sec:diffuse}

Electrons accelerated at each merger event lose energy through ICS on the
photons of the cosmic microwave background, that are upscattered to higher 
energies, up to gamma ray energies. At each merger, two main shock surfaces 
are generated \cite{gb2002}, each one able to accelerate its own population 
of non thermal electrons. The balance between injection and energy losses drives the electrons toward their time independent equilibrium distribution, with a spectrum one power steeper than the injection spectrum.
The rate of gamma ray production from each merger, $Q_\gamma(E_\gamma,M_1,M_2)$ (see \cite{blasi 2001} for details of the calculation of $Q_\gamma$ from the
electron spectrum) is thus the sum of the contributions from the two shocks, 
which in general depends upon the masses $M_1$ and $M_2$ of the merging structures. 
Each shock generates a power law spectrum of gamma rays with two different 
slopes (determined by the compression factors at the shocks as explained 
in the previous section) and a cutoff determined by the maximum energy of the 
accelerated electrons. The flux of gamma radiation 
(in units of $\rm phot~ cm^{-2}s^{-1} sr^{-1} GeV^{-1}$) is then given by
$$
I_\gamma(E_\gamma) = \frac{c}{4\pi H_0} \int_0^{z_{max}} dz \frac{1}{S(z)}
\int_0^\infty dM ~n(M,z) \times
$$
\begin{equation}
\int_0^{M} dM' ~{\cal R}(M,M+M',z) Q_\gamma(E_\gamma(1+z),M,M') \Delta t_{mer}(M,M') ,
\end{equation}
where $S(z)=\sqrt{\Omega_m (1+z)^3 + \Omega_\Lambda}$, ${\cal R}(M,M+M',z)$ is 
the merger rate between clusters of masses $M$ and $M'$ at redshift $z$, and 
$\Delta t_{mer}(M,M')$ is the duration of the merger between two clusters with 
given masses $M$ and $M'$, as defined in \cite{gb2002}.

In the literature the expression {\it accretion events} is often used \cite{fujita,sole}
to describe the mergers of small subclusters with a large dark matter halo. This
definition, implying an artificial separation between merger events and accretion, 
may be useful in other contexts, but here may be quite confusing.
As discussed in \S \ref{sec:formation}, the so-called {\it accretion events} 
are the ones that generate high Mach numbers and therefore flat spectra of 
accelerated particles \cite{gb2002}. It would therefore be instructive but 
rather difficult to define the boundary between accretion and merger events. 
In the older literature, the concept of accretion onto a large scale structure 
was discussed in detail (see for instance \cite{bert}) but it had a rather 
different meaning: a small perturbation in the density field grows as a result 
of gravitational instability, so that more matter falls onto the potential well. 
There is a radius, the so-called turnaround radius at which the inflow velocity 
is balanced by the expansion of the universe.
An accretion shock is formed at a position that depends on time in the same way as the 
turnaround radius and as the virial radius of the structure. The shock surface carries 
the information of the virialization of the inner region of the cluster. This type 
of accretion is also called {\it secondary infall}, meaning that matter accretes on a 
potential well which has already been formed. In the following, we adopt this
simple approach and calculate the gamma ray production due to particle acceleration
at the accretion shock, and compare it with the result of the gamma ray production
from mergers.

For simplicity let us assume that the shock is located exactly at the virial 
radius $R_{vir}$. The total energy per unit time flowing across the shock is then:
$$
L_{tot} = \frac{1}{2} \rho_{cr} \frac{\Omega_b}{\Omega_m}(1+z)^3 v_{ff}^3 
4\pi R_{vir}^2,
$$
where $\rho_{cr} \Omega_m$ is the matter density of the universe, $\Omega_B$ is the
baryon fraction and $v_{ff}$ is the free-fall velocity of the matter at the 
distance of the shock radius. The secondary infall just described is a simplification
of the mass flow through large scale shocks in the filamentary structures seen
in N-body simulations. Although the geometry is different, the total energy crossing
the shock per unit time and per unit surface should not be very different from
the same quantity calculated for spherical inflow.

The accretion shock, by definition, propagates in a cold (non-virialized) medium,
therefore its Mach number may be very high, although the typical speed of the 
shock is of the same order of magnitude as the merger speed of two clusters. The 
exact value of the Mach number depends on the temperature of the medium before 
entering the overdense virialized region. If we take for such temperature the range 
of values $T=10^4 - 10^6$ K, and a typical shock speed of $\sim 10^8$ cm/s, the 
corresponding Mach numbers range between 10 and 100. These Mach numbers, being 
much larger than unity, correspond to spectra of accelerated particles which are 
$\sim E^{-2}$, with a cutoff at the maximum energy (Eq. \ref{eq:Emax}). 
If the intergalactic medium were pre-heated before the gravitational collapse,
these Mach numbers could be lower \cite{susumu}.

The rate of gamma ray production from the accelerated electrons, 
$Q_\gamma(E_\gamma,M,z)$, can be calculated in the usual way \cite{blasi 2001}.
Given $Q_\gamma(E_\gamma,M,z)$ (in $\rm phot~s^{-1}~GeV^{-1}$) in an accreting 
cluster of mass $M$ at redshift $z$, the diffuse flux is
\begin{equation}
I(E_\gamma) = \frac{c}{4\pi H_0} \int_0^{z_{max}} ~ dz \frac{1}{S(z)}
\int^{\infty}_0~dM n(M,z) Q_\gamma(E_\gamma (1+z),M,z).
\end{equation}

\begin{figure}[thb]
 \begin{center}
  \mbox{\epsfig{file=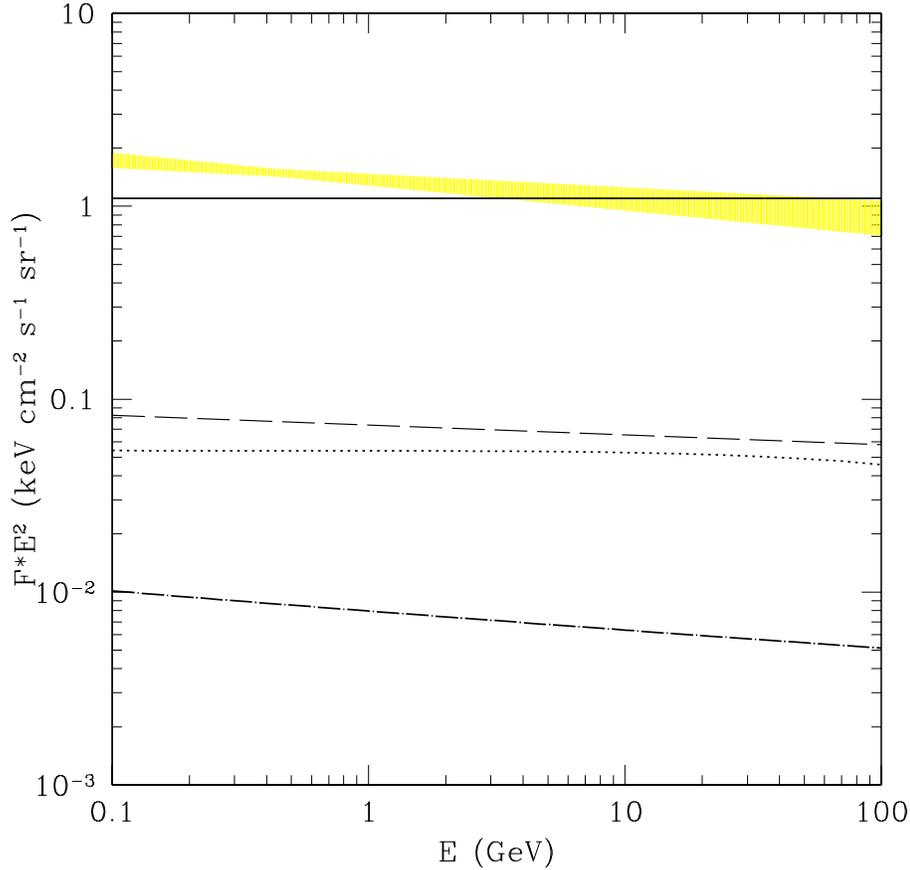,width=13.cm}}
  \caption{\em {Diffuse gamma ray emission from structure formation. The shaded
area is the result EGRET observations. The dashed line is the result of our 
calculations for mergers while the dotted line is the flux of gamma rays from 
accretion. The dash-dotted line assumes a minimum mass of the merging halos
of $10^{13}M_\odot$.}
\label{fig:gamma}}
 \end{center}
\end{figure}
The diffuse flux of gamma radiation from mergers (dashed line) and from accretion 
(dotted line) is plotted in Fig. \ref{fig:gamma}, where an acceleration efficiency 
(for electrons) of $5\%$ is assumed. The observed EDGRB is the shaded region \cite{EGRET}. 
In the same figure we plot for comparison the predictions of Ref. \cite{lw} 
(solid line), where our same acceleration efficiency was adopted. The 
meaning of the dash-dotted line will be explained below. 

Three conclusions are evident: 

1) the flux of gamma radiation from both mergers and accretion is a factor 
$\sim 10$ smaller than the observed EDGRB and smaller than the flux predicted 
in \cite{lw,totani}, by the same factor. An acceleration efficiency of the order
of $50\%$ should be adopted in order to reproduce observations. This would be
unreasonable for electrons as accelerated particles, and would violate our
initial assumption of shock acceleration in the linear regime (no backreaction 
of the accelerated particles on the shock);

2) the gamma ray diffuse flux from mergers is at the same level as that due 
to accretion (secondary infall); 

3) all predicted spectra are approximately power laws with slopes between 
2 and 2.1.

The flatness of the spectrum may give us the key to understand the differences between our
result and that of \cite{lw,totani}. The authors of Ref. \cite{lw,totani} assume 
that all the shocks are strong, so that the spectrum of the accelerated particles
is fixed to $E^{-2}$. In our approach the Mach numbers of the shocks, as well as the 
energy flux through each shock are calculated self-consistently, so that the spectrum 
of the diffuse radiation is the result of the superposition of spectra with different 
slopes. However, although major mergers are energetically very powerful, they generate 
steep spectra of accelerated particles \cite{gb2002}. Therefore the contribution of 
relativistic electrons from these mergers is small compared with that of smaller, 
less energetic mergers, which however produce flatter spectra. This is the reason why 
the gamma ray spectra resulting from mergers have almost the same spectrum as that 
predicted in \cite{lw,totani}. Our absolute normalization is however a factor 
$\sim 10$ lower, which may suggest that merger related shocks are not always strong, and 
in fact they are almost always weak, as illustrated in Fig. \ref{fig:mach}.

\begin{figure}[thb]
 \begin{center}
  \mbox{\epsfig{file=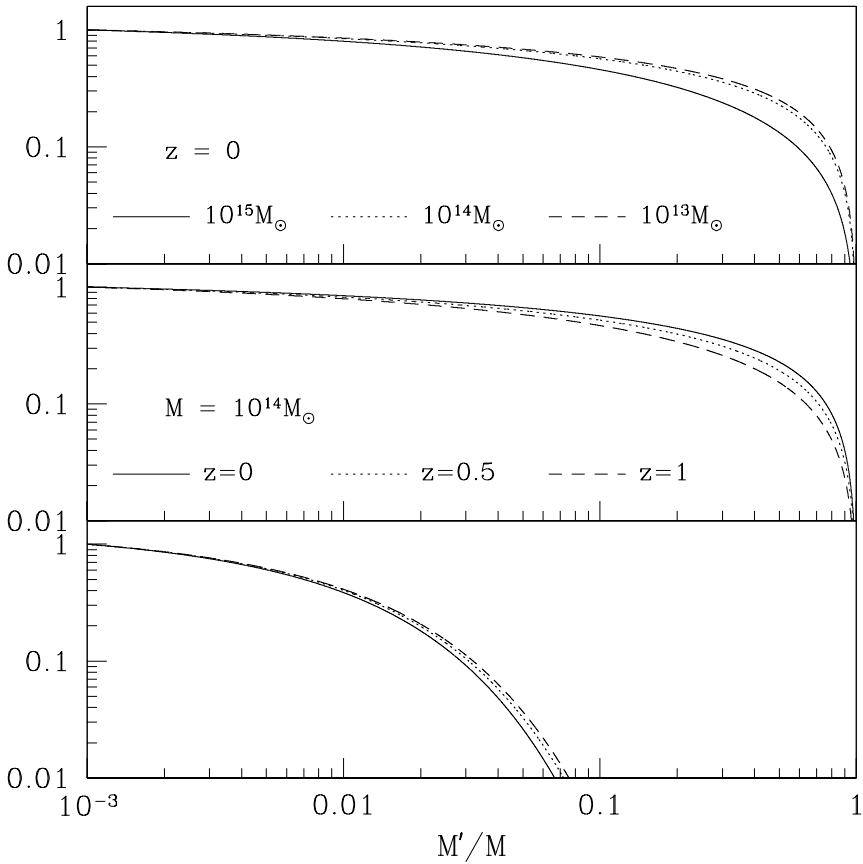,width=13.cm}}
  \caption{\em {Upper panel: normalized energy flux per unit time through the merger 
related shocks of a cluster with mass $10^{15}M_\odot$ (solid line), $10^{14}M_\odot$ 
(dotted line) and $10^{13}M_\odot$ (dashed line) at redshift $z=0$ with clusters with
mass larger than $M'/M$. Middle panel: Same as above for a cluster of mass
$10^{14}M_\odot$ at redshifts $z=0$ (solid line), $z=0.5$ (dotted line) and $z=1$ 
(dashed line). Lower panel: normalized energy flux through the merger related shocks 
that contribute to the diffuse gamma ray background above $100$ MeV, for the same halos
as in the upper panel.}
\label{fig:compare}}
 \end{center}
\end{figure}

Our results can be better clarified by using Fig. \ref{fig:compare}. 
The upper panel shows the average normalized energy flux per unit time through the merger 
related shocks of a cluster with mass $10^{15}M_\odot$ (solid line), $10^{14}M_\odot$ 
(dotted line) and $10^{13}M_\odot$ (dashed line) at redshift $z=0$, as a function of 
the mass of the merging subcluster (here $M^{\prime}/M \leq 1$ is the ratio between the 
masses of the two subclusters). The curves represent the
energy flux contributed by mergers with mass ratio larger than $M^{\prime}/M$. The energy 
flux sums up to $\sim 70-80\%$ of the total for subcluster masses larger than $\sim 0.1 M$. 
This implies that the energy flux that crosses the shocks formed during mergers of 
the cluster with mass $M$ and subclusters with masses smaller than $0.1 M$ is small
($20-30\%$). In other words, the energy flux is dominated by major mergers. 

In the second panel of Fig. \ref{fig:compare} the energy flux is plotted for a cluster 
of fixed mass of $10^{14} M_\odot$ at three redshifts, $z=0$ (solid line), $z=0.5$ 
(dotted line) and $z=1$ (dashed line). The same conclusions explained above hold here. 

The third panel is the most interesting: it represents the normalized energy flux 
through the merger related shocks that contribute to the diffuse gamma ray background 
above $100$ MeV, for clusters of masses as labelled in the upper panel, at $z=0$. 
It is immediately clear that most of the contribution to the gamma ray emission
is provided by mergers with small mass ratios, $M'/M\le 10^{-2}$, namely the 
ones having the larger Mach numbers (see Fig. \ref{fig:mach}). 

Summarizing, while most of the energy flows through shocks associated to major mergers, 
the energy flux that contributes to the gamma ray background is the one that crosses 
strong shocks, occurring when a large cluster encounters a subcluster with 
$\sim 0.01$ times the mass of the larger cluster. This may explain why the diffuse gamma 
ray background as derived in the present paper is substantially lower than that 
estimated in previous calculations \cite{lw,totani}. 
In a recent paper \cite{lwrevised}, a reevaluation of the diffuse gamma ray
background from large scale structure shocks was carried out and there seems to be 
there a closer agreement with the conclusions of our calculations. In \cite{lwrevised}
many issues were discussed as possible reasons for the discrepancy with the 
results of \cite{lw}: one of the points that the authors correctly find out is that simulations do suggest the formation of weak shocks, although difficult to identify. Another numerical calculation was also carried out in Ref. 
\cite{miniati}. 
In \cite{lwrevised} the authors emphasize the difficulties in the identification 
of shocks with Mach number below $10$ (and the impossibility to detect shocks with
Mach numbers below $\sim 3-4$). The Mach number distribution obtained in \cite{lwrevised}
presents an artificial peak at the threshold of detectability (${\cal M}\sim 4$). 
A peak is seen also in \cite{miniatishock}, but it is argued that it is not an artifact
of the numerical procedure. In our semi-analytical approach no peak is 
found at ${\cal M}\sim 4$, while the Mach number distribution seems peaked at 
$\sim 1.5$ \cite{gb2002}.

Another ingredient introduced in our calculation (as well as in \cite{lwrevised}) but not in \cite{lw} is the redshift dependence of the $\gamma-$ray emissivity. This also induces $\gamma-$ray diffuse fluxes smaller than those in \cite{lw}, as also pointed out in \cite{lwrevised}.

The flux of diffuse gamma rays due to accretion (secondary infall) has a spectrum 
which is exactly $E^{-2}$ because the Mach number of the accretion shock is always 
much larger than unity. It is somewhat surprising that the diffuse gamma ray background 
contributed by electrons accelerated at the accretion shock is comparable with that produced 
in merger events (the latter, as stressed above, is dominated by mergers between 
clusters with $M'/M \ll 1$, that in some literature are indeed defined as 
accretion events). In this respect however some additional discussion is required:
for a cluster with mass $10^{14}M_\odot$ a mass ratio $M'/M\sim 10^{-2}$ corresponds
to a substructure with mass $10^{12} M_\odot$, comparable with the mass of our galaxy. 
This clearly does not make physical sense. Galaxies move within the intracluster medium 
without their medium being shocked. It is more likely that a bow shock is formed in 
front of the galaxy, due to the internal pressure of the galactic medium \cite{stevens}. 
Actually simulations show that large galaxies penetrating the intracluster medium of a rich
cluster can even be completely stripped of their gas content \cite{stevens}.
This suggests that a low mass cutoff should be imposed in the calculation of the gamma ray 
diffuse background from cluster mergers. The dash-dotted line in Fig. \ref{fig:gamma} 
has been obtained by considering only structures with virial masses larger than $10^{13}
M_\odot$, corresponding to galaxy groups. The diffuse background of gamma rays is
a factor $\sim 10$ smaller in this case and is slightly steeper in spectrum. This
happens because the main contribution comes from mergers between clusters with 
masses $M_{min}=10^{13} M_\odot$ and $M=M_{min}/10^{-2}\approx 10^{15} M_\odot$.
Clusters with masses as large as $10^{15} M_\odot$ are already on the tail 
of the Press-Shechter distribution even at $z=0$, therefore the corresponding contribution to
the diffuse background is suppressed. On this basis, the dash-dotted line in Fig. 
\ref{fig:gamma} is the most realistic estimate of merger shocks to the diffuse 
gamma ray background, amounting to $\sim 1\%$ of the observed EDGRB (this result 
agrees with the estimate in \cite{dermer}). On the other hand the strong shocks 
associated to accretion of matter onto a cluster may generate a gamma ray background
as large as that plotted as a dotted line in Fig. \ref{fig:gamma}, and this contribution 
remains at the level of $\sim 10\%$ of the observed EDGRB. This may be considered 
as the most realistic prediction of the contribution of clusters of galaxies 
to the EDGRB.

Our results are in agreement with the recent estimate of the diffuse gamma ray flux 
from rich clusters carried out by cross correlating high galactic latitude EGRET data 
with the location of Abell clusters \cite{scharf}.

\section{Conclusions}\label{sec:conclusions}

We calculated the contribution of structure formation to the EDGRB radiation. We find that 
this contribution is only $1-10\%$ of the observed flux of the alleged extragalactic radiation 
above $100$ MeV as measured by EGRET. The calculation has been carried out in two different
scenarios, one that relies upon the hierarchical scenario for structure formation, and
the other that is based on the secondary infall (accretion) of matter onto a potential 
well which has already been formed. 

In the hierarchical approach, large structures are formed by merging of smaller halos 
whose mass is dominated by dark matter. The baryon components of these halos, moving 
supersonically, develop shock surfaces that in principle can accelerate particles to
relativistic energies. We evaluate the Mach numbers of these shocks following the 
recipe introduced in \cite{gb2002,Takizawa99}, that allows us to self-consistently 
calculate the spectra of accelerated particles at each merger event. This 
distinguishes our approach from previous calculations \cite{lw,totani} where the shocks 
were all assumed strong (infinite Mach number), so that the spectra of accelerated particles 
were by definition $E^{-2}$. We find that this assumption may lead to incorrect conclusions,
as evidenced by the comparison between our results and those in \cite{totani}.

In \cite{gb2002} we investigated the role of protons first accelerated and then 
diffusively confined in large scale structures \cite{bbp97}. We find that the
spectral shape of the protons is very steep and is unlikely to produce a relevant
effect on high energy radiation generated by clusters of galaxies. This conclusion
is mostly due to the fact that major mergers, that energetically dominate over
smaller merger events, generate weak shocks and therefore steep particle spectra.

In the present paper we focused our attention on electrons as accelerated
particles. Their ICS energy losses were in fact proposed \cite{lw,totani} as 
responsible for upscattering the photons of the cosmic microwave background to
gamma ray energies, therefore generating a diffuse background of gamma radiation
accompanying the process of structure formation. While previous calculations
suggest that the observed EDGRB may be saturated by the contribution of particles
accelerated during structure formation, we find here that at most $10\%$ of the
observed background can in fact be explained in this way. More recent numerical 
calculations \cite{lwrevised,miniati} seem to lower previous predictions. 

Although structure formation is generally believed to follow the hierarchical 
picture outlined above, some secondary infall of matter onto forming structures
must occur, and is in fact observed in numerical N-body simulations in the form
of filamentary-like accretion flows. In order to account for this contribution 
we adopt a simple model, similar to that proposed in \cite{bert}, in which an
accretion shock is formed at approximately the virial radius of a structure that
is accumulating matter from the expanding universe. This shock, as those formed
in the filaments, propagates in the cold unshocked medium and may have very 
large Mach numbers, of order 10-100 (possibly lower if pre-heating occurs \cite{susumu}).
Particles accelerated at this type of shocks have the flattest spectrum allowed 
in the linear theory of shock acceleration, namely $E^{-2}$. The diffuse gamma
ray background due to ICS of electrons accelerated at accretion shocks is also
of order $10\%$ of the observed EDGRB, and may be larger that that due to mergers. 
The reason for this result is the following: although the energy 
flux through merger shocks is larger than that crossing accretion shocks, the 
former is mainly converted into very steep spectra (due to the weakness of the shocks),
while the latter is more easily channelled into particles that may contribute
to gamma rays with energy above $\sim 100$ MeV, because of the flatter spectra.

In the perspective of future work in this field, it seems to us that priority
should be given to improve N-body simulations in order to have a better handling 
of the shocks with intermediate strength generated during structure formation. 
This will allow a self-consistent treatment of both the gas heating and the 
acceleration of particles at these shocks, and make a solid case in favor or against
clusters of galaxies as sources of high energy gamma radiation. 

The results presented in this paper suggest that clusters of galaxies, and
more in general structure formation, do not contribute appreciably 
to the EDGRB. However, several clusters could be observed as single gamma 
ray sources by future experiments such as GLAST (Blasi and Gabici, in 
preparation) and provide useful information on the non thermal history 
of these large scale structures.

{\bf Acknowledgments:} We are grateful to U. Keshet, C. Dermer and F. Stecker for useful comments.

\end{document}